\documentclass[prl,aps,showpacs,twocolumn]{revtex4}
\usepackage{bm}
\usepackage{graphicx}
\usepackage{amsbsy}
\usepackage{amsmath}
\usepackage{amsfonts}

\newcommand{\bra}[1]{\langle{#1}|}
\newcommand{\ket}[1]{|{#1}\rangle}
\newcommand{\nn}{\nonumber}
\newcommand{\dg}{^\dagger}

\begin{document}

\title{On improving single photon sources via linear optics and photodetection}

\author{ Dominic W. Berry$^1$, Stefan Scheel$^2$, Barry
C. Sanders$^{1,3}$, and Peter L. Knight$^2$} 
\affiliation{${}^1$ Australian Centre for Quantum Computer Technology,
Department of Physics, Macquarie University, Sydney, New South Wales
2109, Australia \\
$^2$ Quantum Optics and Laser Science, Blackett Laboratory, 
Imperial College London, Prince Consort Road, London SW7 2BW, UK \\ 
$^3$ Quantum Information Science Group, Department of Physics and
Astronomy, University of Calgary, Alberta T2N 1N4, Canada}

\begin{abstract}
In practice, single photons are generated as a mixture
of vacuum with a single photon with weights $1-p$ and
$p$, respectively; here we are concerned with increasing $p$
by directing multiple copies of the single photon-vacuum mixture into
a linear optic device and applying photodetection on some
outputs to conditionally prepare single photon states
with larger $p$. We prove that it is impossible, under certain
conditions, to increase $p$ via linear optics and conditional
preparation based on photodetection, and  we also establish a class of
photodetection events for which $p$ can be improved. In addition we
prove that it is not possible to obtain perfect ($p=1$) single photon
states via this method from imperfect ($p<1$) inputs.
\end{abstract}

\date{\today}


\maketitle

Single photon sources are important, for example to achieve
secure quantum key distribution \cite{key} and for linear optic
quantum computation \cite{LOQC}, yet generating single photons remains
challenging. The traditional method involves photodetection
on one output mode from a nondegenerate parametric downconversion
process to post-select a single photon in the correlated mode \cite{down}.
More recently alternative single-photon sources have been employed,
including molecules \cite{mol}, quantum wells \cite{well},
color centers \cite{col}, ions \cite{ion} and quantum dots \cite{dot}.
In these examples, the 
probability of more than one photon being produced is much
lower than that for a Poissonian process, but the vacuum
contribution can be quite high. Under ideal conditions,
with a single mode output stable over time, the output
state can be described by the density matrix
\begin{equation}
\label{eq:initialstate}
\hat{\rho}_p = (1-p) \ket 0 \bra 0 +
p \ket 1 \bra 1 ,
\end{equation}
with $p$ the probability of obtaining a single photon;
$p$ is sometimes referred to as the efficiency for
producing single photons. Whereas the theoretical 
single-photon state corresponds to $p=1$, the experimental
single-photon state is a mixture of the single photon
with the vacuum.

Increasing the value of $p$ (the efficiency of producing
single photons) is an increasingly important objective
because of requirements for quantum optics experiments,
especially those concerned with quantum information processing.
Much of this effort is directed to improving sources,
but here we pose the question as to whether 
a source of single photons with efficiency $p$ can be
processed via linear optics and photodetection to
yield fewer photons but with higher $p$. More specifically,
would it be possible to direct $N$ copies of these
single photon states with efficiency $p$ into an $N$-channel
passive interferometer (an interferometer that involves
only passive, linear optical elements) to yield an output
single photon with efficiency $p'>p$ for certain photodetection
records on the other outputs? If the answer is yes,
then interferometry and photodetection could be employed 
to improve the efficiency of single-photon sources.

In fact the answer will be no, provided we consider detection results
where all but one of the photons are detected. This is the most
straightforward way of ensuring that the output state contains at most
one photon. This restriction is necessary in most applications of linear
optics quantum information processing since two-photon contributions can
have unwanted side effects, e.g.\ allowing for certain multiphoton attacks
in quantum key distribution applications. If we allow other
detection results, it is possible for low-efficiency (small $p$)
single-photon states to yield, via linear optics and conditional
preparation based on photodetection, an output with a larger probability
for a single photon. However, these schemes also yield non-zero
probabilities for photon numbers above one.

A passive interferometer is comprised of beam splitters,
mirrors, and phase shifters. Each of these elements preserves
total photon number from input to output under ideal conditions.
No energy is required to operate these optical elements,
hence the term passive. These are also known as linear optical
elements. More generally polarization transforming elements
can be included, but here we are concerned only with a scalar
field treatment; in fact polarization effects could be included
by doubling the number of channels and treating the two 
polarizations in a mode as two separate channels.
Mathematically, a passive interferometer transforms the amplitude
operators of the incoming fields $\hat{\bm{a}}$ via the matrix
transformation
$\hat{\bm{a}}^\dagger\mapsto\bm{\Lambda}^T\hat{\bm{a}}^\dagger$ with 
$\Lambda\in \text{U}(N)$ \cite{VogelWelsch}.

\begin{figure}[ht]
\centerline{\includegraphics[width=4cm]{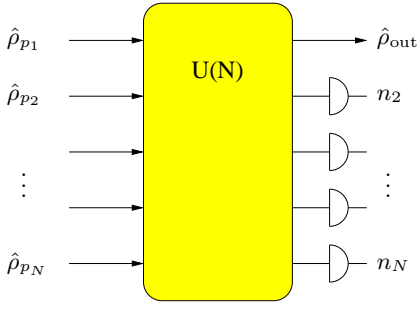}}
\begin{picture}(0,0)
\put(-80,108){$\hat{\rho}_{p_1}$}
\put(-80,87){$\hat{\rho}_{p_2}$}
\put(-75,50){$\vdots$}
\put(-80,23){$\hat{\rho}_{p_N}$}
\put(60,108){$\hat{\rho}_{\text{out}}$}
\put(60,87){$n_2$}
\put(63,50){$\vdots$}
\put(60,23){$n_N$}
\end{picture}
\caption{\label{fig:scheme} Schematic setup of the network. We assume
$N$ incoming modes prepared in the state \eqref{eq:input} with
different $p_i$.}
\end{figure}

In the general case we start with a supply of $N$ mixed states of the form
\eqref{eq:initialstate}. For additional generality we allow the
different inputs to have different probabilities for finding a single photon,
$p_i$, and we denote the maximum of these probabilities by $p_{\rm max}$.
The initial state may be expressed as
\begin{align}
\label{eq:input}
\hat{\rho}_{\text{in}}^{(N)} & = \bigotimes_{i=1}^N \big[ 
(1-p_i) \ket 0 \bra 0 + p_i \ket 1 \bra 1 \big] \nn \\
& = \sum_{\bm{s}} P_{\bm{s}}
\left( \prod_i (\hat a_i\dg)^{s_i} \ket{0} \otimes \text{h.c.} \right),
\end{align}
where
$P_{\bm{s}}$ $\!=$ $\!\prod_i p_i^{s_i}(1-p_i)^{1-s_i}$,
and the vector $\bm{s}$ $\!=$ $\!(s_1, \cdots , s_N)^T$, ($s_i=0,1$), gives
the photon numbers in the inputs. The interferometer transformation
$\hat{\bm{a}}^\dagger\mapsto\bm{\Lambda}^T\hat{\bm{a}}^\dagger$
results in the output state 
\begin{equation}
\hat\rho_{\rm trans}^{(N)} = \sum_{\bm{s}} P_{\bm{s}}
\left[ \prod_i \left(\sum_k \Lambda_{ki} \hat a_k\dg\right)^{s_i} \ket{0}
\otimes \text{h.c.} \right].
\end{equation}

Without loss of generality, we take mode 1 to be the mode in which we
wish to improve the photon statistics. We perform photodetections on
the other $N-1$ modes, and examine the final state in mode 1
conditioned on the results of these photodetections. The total number
of photons detected is $D$, and the number of photons detected in mode
$j$ is $n_j$ (see Fig.~\ref{fig:scheme}). It is easy to see that no
better result can be obtained by performing photodetections on fewer
than $N-1$  modes; this would be equivalent to averaging over the
photocounts for some of the modes. We assume ideal photodetection in
this analysis in order to determine the best results possible using
linear optics and photodetection. 

The conditional state in mode 1 after photodetection in modes 2 to $N$ is
\begin{equation}
\hat\rho_{\rm out}^{(N)} = \sum_{n_1=0}^N c_{n_1} \ket{n_1}\bra{n_1},
\end{equation}
where $n_1$ is the photon number in the remaining output mode (rather than the
number of photons detected at a photodetector).
Each coefficient $c_{n_1}$ is given by
\begin{equation}
\label{expect}
c_{n_1} = K\bra{\bm{n}}\hat\rho_{\rm trans}^{(N)}\ket{\bm{n}},
\end{equation}
where $\ket{\bm{n}}$ is a tensor product of number states in each of the
output modes. The normalization constant $K$ equals
\begin{equation}
K=\left[\sum_{n_1=0}^N \bra{\bm{n}}\hat\rho_{\rm trans}^{(N)}
\ket{\bm{n}}\right]^{-1}. 
\end{equation}
In order to find the expectation value in Eq.\ \eqref{expect}, we
first introduce some 
notation. Let $\Phi_{\bm{s}}=\{ i | s_i = 1 \}$,
$\Sigma_{\bm{s}}=\sum_i s_i$  (so $\Sigma_{\bm{s}}$ is the number of
elements in $\Phi_{\bm{s}}$),
and let ${\rm Y}_{\bm{s}}$ be the set that consists of all vectors
comprised of the elements of $\Phi_{\bm{s}}$.
In addition we use the notation
$K'=K/\prod_{j=2}^N n_j!$ and $\Sigma_{\bm{n}} = \sum_j n_j$.
Using this notation, $c_1$ is given by
\begin{equation}
\label{general}
c_1 = \frac {K'}{n_1!} \sum_{\bm{s};\Sigma_{\bm{s}}=\Sigma_{\bm{n}}}
P_{\bm{s}}\left| S_{\bm{s},\bm{n}}\right|^2,
\end{equation}
where
\begin{align}
S_{\bm{s},\bm{n}} &
= \sum_{\bm{\sigma}\in {\rm Y}_{\bm{s}}}
\left( \Lambda_{1,\sigma_1} \cdots 
\Lambda_{1,\sigma_{n_1}} \right)
\nn \\ &
\cdots \left(
\Lambda_{N,\sigma_{\Sigma_{\bm{n}}-n_N+1}} \cdots 
\Lambda_{N,\sigma_{\Sigma_{\bm{n}}}} \right).
\end{align}

In order to determine if there is an improvement in the probability of
finding a single photon, we need to determine the value of $c_1$. However,
determining $c_1$ requires evaluating $K$, which requires evaluating
the expectation values in Eq.\ \eqref{expect} for all possible values
of $n_1$. Instead, we will consider the ratio $c_1/c_0$. There are three
main advantages to considering this quantity: \\
1. The common constant $K'$ cancels, so this expression is easily
evaluated analytically. \\
2. If $c_1/c_0$ is not greater than $R\equiv p_{\rm max}/(1-p_{\rm max})$,
then it is clear that $c_1\le p_{\rm max}$. Thus we can determine
those cases where there is {\it no} improvement. \\
3. For $p_{\rm max}\ll 1$, $c_0\approx 1$ and $R\approx p_{\rm max}$.
Therefore the improvement in $c_1/c_0$ over $R$ is approximately the
same as the improvement in the probability of a single photon over
$p_{\rm max}$. \\

Ideally, we would determine the interferometer and detection pattern
such that $c_1/c_0$ is maximized, but this does not appear to be
possible analytically. However, we can place an upper limit on $c_1/c_0$
in the following way. Let us express the summation for $c_0$ as
\begin{equation}
\label{altno}
c_0 = \frac {K'}{N-D} \sum_{\bm{s}; \Sigma_{\bm{s}}=D+1}
\sum_{k;s_k=1} P_{\bm{s}^k} \left| S_{\bm{s}^k,\bm{n}'} \right|^2,
\end{equation}
where $s^k_i=s_i$ except for $s^k_k=0$, and $n'_j=n_j$ except $n'_1=0$.
The quotient of $N-D$ takes account of a redundancy
in the sum. Each alternative input $\bm{s}^k$ has $N-D$
zeros, so there are $N-D$ possible alternative $\bm{s}$
that give the same $\bm{s}^k$.
We may reduce this quotient slightly if we take account of
the possibility that some of the inputs have zero photon
probability. Let there be $N-M$ inputs with $p_i=0$, so the 
maximum total number of photons is $M$. If we limit the first sum
in Eq.\ \eqref{altno} to $\bm{s}$ such that $P_{\bm{s}}\ne 0$, then
the redundancy is $M-D$. Therefore we obtain 
\begin{align}
c_0 &= \frac {K'}{M-D} \sum_{\substack{\bm{s}; P_{\bm{s}}\ne 0 \\
\Sigma_{\bm{s}}=D+1 }} \sum_{k;s_k=1} P_{\bm{s}^k} \left| S_k \right|^2 \nn \\
&= \frac {K'}{M-D} \sum_{\substack{\bm{s}; P_{\bm{s}}\ne 0 \\
\Sigma_{\bm{s}}=D+1 }} P_{\bm{s}} \sum_{k;s_k=1} \frac{1-p_k}{p_k} \left| S_k \right|^2,
\end{align}
where $S_k = S_{\bm{s}^k,\bm{n}'}$.
Since we have limited the sum to terms where $P_{\bm{s}} \ne 0$,
$p_k$ is nonzero, and thus the ratio $(1-p_k)/p_k$ does not
diverge. Since $p_k$ does not exceed $p_{\rm max}$, we have the
inequality
\begin{equation}
\label{ineq0}
c_0 \ge \frac {K'}{M-D}\frac{1-p_{\rm max}}{p_{\rm max}} \sum_{\bm{s};
\Sigma_{\bm{s}}=D+1} P_{\bm{s}} \sum_{k;s_k=1} \left| S_k \right|^2.
\end{equation}
Here we are able to omit the condition $P_{\bm{s}} \ne 0$ because terms
with $P_{\bm{s}}= 0$ are zero anyway.
We may re-express the equation for $c_1$ as
\begin{equation}
\label{onephoton}
c_1 = K' \sum_{\bm{s};\Sigma_{\bm{s}}=D+1} P_{\bm{s}}
\left| \sum_{k;s_k=1} \Lambda_{1k} S_k \right|^2.
\end{equation}
We therefore obtain
\begin{equation}
\label{ineq1}
c_1 \le K' \sum_{\bm{s};\Sigma_{\bm{s}}=D+1  }
 P_{\bm{s}} \sum_{k;s_k=1} \left| S_k \right|^2.
\end{equation}
Combining Eqs.\ \eqref{ineq0} and \eqref{ineq1} gives
\begin{equation}
\label{ineq1a}
\frac{c_1}{c_0}\le (M-D)\frac{p_{\rm max}}{1-p_{\rm max}}.
\end{equation}

This yields an upper limit on the ratio between the one and zero
photon probabilities. One application of this result is that it is
impossible to get one photon with unit probability, as it would
be necessary for this ratio is infinite.
Another consequence of (\ref{ineq1a}) is that for $D=M-1$ (i.e.\ the
number of photons detected one less the maximum input number) an
improvement can never be achieved. This case is important
because it is the most straightforward way of eliminating the
possibility of two or more photons in the output mode.

In the remainder of this article, we will investigate situations in
which the single-photon contribution can be enhanced.
As $M\le N$, and $D\ge 0$, the upper limit on the improvement
in $c_1/c_0$ is simply $N$. This is also the upper limit in how
far $c_1$ can be increased above $p_{\rm max}$. We will now consider a
scheme that gives a linear improvement in $c_1/c_0$, though not as high
as $N$. In order to obtain a large value for the ratio $c_1/c_0$, we want
the inequality in Eq.\ \eqref{ineq1} to be as close to equality as
possible. In turn, this means that we want the vectors $(\Lambda_{1k})$ and
$(S_k)$ to be as close to parallel as possible.
For this, we consider the interferometer given by
\begin{align}
\label{interfer}
\Lambda_{21} =-\epsilon, & \quad \Lambda_{22} = \sqrt{1-\epsilon}, \nn \\
\Lambda_{i1} =\sqrt{(1-\epsilon^2)/(N-1)}, & \quad
\Lambda_{i2}=\epsilon/\sqrt{N-1}, 
\end{align}
for $i\ne 2$ (the values of $\Lambda_{ij}$ for $j>2$ do not enter into
the analysis). Here $\epsilon$ is a small number, and we will ignore
terms of order $\epsilon$ or higher. Now let $p_i=p_{\rm max}$, and
consider the measurement record where zero photons are detected in modes
3 to $N$, and $D$ photons are detected in output mode 2.
To determine $c_{n_1}$, note first that $\Lambda_{22}\gg \Lambda_{2i}$
for $i\ne 2$, so we may ignore those terms in the sum for
$S_{\bm{s},\bm{n}}$ where $\Lambda_{22}$ does not appear. Each term has
magnitude $\Lambda_{11}^{n_1}\Lambda_{22}\Lambda_{21}^{D-1}$
\footnote{Here we are using $\Lambda_{21}$ and $\Lambda_{11}$ to indicate
the values of $\Lambda_{2i}$ and $\Lambda_{1i}$ for $i\ne 2$.}, and there
are $D(D+n_1-1)!$ such terms. Therefore, provided $s_2=1$,
\begin{equation}
S_{\bm{s},\bm{n}} \approx
D(D+n_1-1)!\Lambda_{11}^{n_1}\Lambda_{22}\Lambda_{21}^{D-1}
\,,\quad \epsilon\ll 1. 
\end{equation}
In the summation for $c_{n_1}$, we have $\binom{N - 1}{D-n_1-1}$
different combinations of inputs such that $\Sigma_{\bm{s}}=D+n_1$ and $s_2=1$.
Combining these results, we have
\begin{align}
\label{total}
c_{n_1} &\approx
\frac{K'}{n_1!}p_{\rm max}^{D+s_1}(1-p_{\rm max})^{N-D-s_1} \nn \\ & \times 
\frac{(N-1)!D^2(D+n_1-1)!}%
{(N-D-n_1)!}\Lambda_{11}^{2n_1}\Lambda_{22}^2\Lambda_{21}^{2D-2}
\nn \\ 
&\approx K'' \left(\frac{R}{N-1}\right)^{n_1}
\frac{(D+n_1-1)!}{n_1!(N-D-n_1)!}, 
\end{align}
We have combined those factors that do not depend on $n_1$ into a new
constant $K''$, and used $\Lambda_{11}\approx 1/\sqrt{N-1}$.

Using Eq.\ \eqref{total} we find that
\begin{equation}
\frac{c_1}{c_0} \approx R \frac{D(N-D)}{N-1} \,,\quad \epsilon\ll 1.
\end{equation}
The maximum improvement in the ratio $c_1/c_0$ is obtained for
$D=\lceil N/2 \rceil$, where $c_1/c_0=R\lfloor N^2/4 \rfloor/(N-1)$.
The multiplying factor $\lfloor N^2/4 \rfloor/(N-1)$ is larger than 1 for
all $N\ge 4$. Thus we find that, provided there are at least 4 modes, we
may obtain an improvement in the ratio $c_1/c_0$. For $p_{\rm max}\ll 1$,
$c_1 \approx p_{\rm max}\lfloor N^2/4 \rfloor/(N-1)$. For large $N$, the
probability of a single photon increases approximately as $N/4$. This is
linear with $N$, but is not quite as large as the upper limit of $N$.

However, there are some drawbacks to this interferometry scheme [i.e.\ using
the interferometer \eqref{interfer}]. The first
drawback is the two-photon contribution. We find that
\begin{equation}
\frac{c_2/c_1}{c_1/c_0} = \frac{(D+1)(N-D-1)}{2D(N-D)}\,,
\end{equation}
which, for $D=\lceil N/2 \rceil$, is less than $1/2$, but it is close to
$1/2$ for large $N$. Since this is the same ratio as for a Poisson
distribution, this is equivalent to using a coherent state.
The two-photon contribution can be reduced by using larger $D$, but this
is at the expense of reducing $c_1$.

The second drawback is that improvements are only obtained for small
$p_{\rm max}$. Although the improvement in the ratio $c_1/c_0$ is
independent of $p_{\rm max}$, improvements in $c_1$ can only be obtained
for values of $p_{\rm max}$ below $1/2$. That is, this method can
only be used to obtain improvements in the probability of a single
photon up to $1/2$, but not to make the probability of a single
photon arbitrarily close to 1.
Note that the above method only gives $c_1>p_{\rm max}$ for four or more
modes. We will now show that it is impossible to obtain an improvement
in the probability of a single photon with fewer than four modes, and
for various combinations of detections with larger numbers of modes.

We first examine the case $D=0$.
Then we have only one term in the sum for $c_0$, and
$c_0 = K' P_{\bm{0}}$. The expression for $c_1$ becomes
\begin{align}
c_1 & = K' \sum_{k=1}^N \frac {p_k}{1-p_k} P_{\bm{0}} \left| \Lambda_{1k}
\right| ^2 
\le K' \frac {p_{\rm max}}{1-p_{\rm max}} \sum_{k=1}^N
P_{\bm{0}} \left| \Lambda_{1k} \right| ^2 \nn \\
& = K' R P_{\bm{0}}
= c_0 R.
\end{align}
Thus we have shown that $c_1/c_0 \le R$, so $c_1 \le p_{\rm max}$.
Hence there can be no improvement in the photon statistics if zero
photons are detected.
We can also obtain a similar result for the case $D=1$,
provided all the input $p_i$ are equal. In that case, we have 
\begin{equation}
c_0 = K' \sum_{k} \frac{p_{\rm max}}{1-p_{\rm max}} P_{\bm{0}}
\left| \Lambda_{2k} \right|^2 = K' R P_{\bm{0}}.
\end{equation}
The value of $c_1$ is given by
\begin{align}
c_1 &= \frac 12 K' \sum_{k}\sum_{l;l\ne k}R^2 P_{\bm{0}}
\left| \Lambda_{1l}\Lambda_{2k} +\Lambda_{1k}\Lambda_{2l} \right|^2 \nn \\
& \le \frac 12 K' R^2 P_{\bm{0}}
\sum_{k,l}
\left| \Lambda_{1l}\Lambda_{2k} + \Lambda_{1k}\Lambda_{2l} \right|^2
= K' R^2 P_{\bm{0}}.
\end{align}
In the last line we have used the fact that $\Lambda_{1k}$ and
$\Lambda_{2k}$ are orthonormal. Thus we again find $c_1/c_0\le R$, so
$c_1 \le p_{\rm max}$.

These results clearly eliminate the possibility of improving the
probability of finding one photon with a two-mode interferometer. We
have shown that detecting zero photons does not give an improvement,
and if one photon is detected, then we must have $M-D=1$ or 0, so
there again can be no improvement. Along the same lines we can also
eliminate the three-mode interferometer. 

We have shown that it is impossible to improve the efficiency of a
single-photon source by channeling more than one low-efficiency
single-photon state into a linear optic interferometer and detecting
all but one of the photons. This eliminates the most straightforward
scheme for obtaining an output state with no more than one photon.
It is possible to obtain an improvement for more general detection
results, but at the expense of non-zero probabilities for two or more
photons. We have not proven that it is impossible to obtain an improvement
in the probability of a single photon while restricting to zero
probability for two or more photons; however, numerical searches
indicate that it is unlikely.

This work was funded in parts by the UK Engineering and Physical
Sciences Research Council. One of the authors (SS)
enjoys a Feodor-Lynen fellowship of the Alexander~von~Humboldt
foundation. BCS appreciates valuable discussions with R.\ Laflamme in 
the early stages of this work. This research has also been supported
by an Australian Department of Education Science and Training
Innovation Access Program Grant to support collaboration in the
European Fifth Framework project QUPRODIS, and by Alberta's
informatics Circle of Research Excellence (iCORE).

\end{document}